\documentclass[reprint,aps,prl,superscriptaddress,nofootinbib]{revtex4-1}

\pdfoutput=1
\usepackage{graphicx}
\usepackage{subfigure}
\usepackage{float}
\usepackage{longtable}

\usepackage{dcolumn}
\usepackage{bm}
\usepackage{amssymb}
\usepackage{latexsym}
\usepackage{booktabs}
\usepackage{amsmath}
\usepackage{multirow}
\usepackage{url}
\usepackage{changes}
\usepackage{soul}
\usepackage{color, xcolor}
\usepackage[colorlinks=true, linkcolor=red, citecolor=blue]{hyperref}

\usepackage[normalem]{ulem}
\usepackage{array}
\usepackage{enumerate}

\def\be{\begin{equation}}
\def\ee{\end{equation}}

\def\bea{\begin{eqnarray}}
\def\eea{\end{eqnarray}}

\newcommand{\pccc}{$\,\rm{pc}\,\rm{cm}^{-3}\,$}
\newcommand{\kms}{km\,s$^{-1}$\, Mpc$^{-1}$}
\newcommand{\ob}{$\Omega_{\rm b}~$}

\newcommand{\apjl}{Astrophys.\ J.\ Lett.}
\newcommand{\aj}{Astron.\ J.}
\newcommand{\mnras}{Mon.\ Not.\ R.\ Astron.\ Soc.}

\newcommand{\natastro}{Nat.\ Astron.}

\begin{document}

\title{Cosmic baryon census with fast radio bursts and gravitational waves}

\author{Ji-Guo Zhang}
\affiliation{Liaoning Key Laboratory of Cosmology and Astrophysics, College of Sciences, Northeastern University, Shenyang 110819, China}

\author{Ji-Yu Song}
\affiliation{Liaoning Key Laboratory of Cosmology and Astrophysics, College of Sciences, Northeastern University, Shenyang 110819, China}

\author{Wan-Peng Sun}
\affiliation{Liaoning Key Laboratory of Cosmology and Astrophysics, College of Sciences, Northeastern University, Shenyang 110819, China}

\author{Ze-Wei Zhao}
\affiliation{Department of Biomedical Engineering, School of Medical Devices, Shenyang Pharmaceutical University, Shenyang 110016, China}

\author{Jing-Fei Zhang}
\affiliation{Liaoning Key Laboratory of Cosmology and Astrophysics, College of Sciences, Northeastern University, Shenyang 110819, China}

\author{Xin Zhang}\thanks{Corresponding author: zhangxin@neu.edu.cn}
\affiliation{Liaoning Key Laboratory of Cosmology and Astrophysics, College of Sciences, Northeastern University, Shenyang 110819, China}
\affiliation{MOE Key Laboratory of Data Analytics and Optimization for Smart Industry, Northeastern University, Shenyang 110819, China}
\affiliation{National Frontiers Science Center for Industrial Intelligence and Systems Optimization, Northeastern University, Shenyang 110819, China}

\begin{abstract}
The cosmic baryon density fraction ($\Omega_{\rm b}$) is intrinsically correlated with the Hubble constant ($H_0$) through the critical density of the Universe. In the context of the decade-long $H_0$ tension, the significant discrepancy between early- and late-Universe measurements of $H_0$ implies that fixing its value or imposing an external prior could bias the baryon census. To address this concern, we construct a late-Universe probe framework that unifies fast radio bursts (FRBs) and gravitational-wave (GW) standard sirens, which can respectively resolve the ``missing baryon'' problem and the $H_0$ tension through their dispersion measures (DMs) and absolute luminosity distances. By combining $104$ localized FRBs with $47$ GW events, we obtain an $H_0$-free measurement of $\Omega_{\rm b}=0.0488\pm0.0064$ ($1\sigma$), in concordance with early-Universe observations of CMB + BBN. The result is tightly anchored by GW-inferred $H_0$ through the strong $\Omega_{\rm b}$-$H_0$ degeneracy. Although the current precision ($\sim 13\%$) is limited by sample size, the growing detections of both FRBs and GWs will make their synergy a powerful probe of low-redshift cosmology.
\end{abstract}

\maketitle

\paragraph{Introduction.}

The concordance cosmology predicts consistency among independent probes across all epochs \cite{Ostriker:1995su}. Yet, while the cosmic microwave background (CMB) and Big Bang nucleosynthesis (BBN) provide precise estimates of the baryon content in the early Universe \cite{Cooke:2017cwo}, about $30\%$ baryons remain undetected in the local Universe \cite{Shull:2011aa}. These missing baryons are thought to reside in diffuse ionized gas that is too faint to be directly observed, known as the ``missing baryon problem'' \cite{Fukugita:1997bi,cen1999baryons}.
Fast radio bursts (FRBs) are bright, millisecond-duration radio transients of extragalactic origin \cite{lorimer2007,thornton2013pop}, which can be broadly classified into repeaters and non-repeaters \cite{2016Natur.531..202S,Sun2025a,Sun2025b}. Their observed dispersion measures (DMs) trace the integrated column density of free electrons along the line of sight (LoS), making them sensitive to baryons in the intergalactic medium (IGM) \cite{mcquinn2013locating}. 
Localized FRBs with measured redshifts thus provide a direct probe of the missing baryons \cite{Deng:2013aga,connor2025gas}. Using 5 and 22 localized FRBs, Refs.~\cite{Macquart:2020lln,Yang:2022ftm} (hereafter Macquart20 and Yang22) inferred $\Omega_{\rm b}$ values consistent with CMB + BBN, suggesting a nearly complete baryon census. With the high event rate and increasing number of localized events, FRBs have emerged as a promising cosmological probe to, e.g., probe dark energy \cite{Zhou:2014yta,Gao:2014iva,Walters:2017afr,dai2023ska,Zhang:2023gye}, the Hubble constant ($H_0$) \cite{Li:2017mek,Hagstotz:2021jzu,Wu:2021jyk,James:2022dcx,Liu:2022bmn,Zhao:2022yiv,Kalita:2024xae,Fortunato:2024hfm,Zhang:2024rra,Wang:2025ugc,Gao:2025fcr,Zhang:2025thh,Gao:2025dls,Zhuge:2025urk,Xu:2025ddk,Jia:2025kvp,Wu:2026mox}, as well as baryonic feedback \cite{Sharma:2025uij,Reischke:2025srr}, and the large-scale structure \cite{Hussaini:2025leu,Wang:2025ibv}; see Refs. \cite{Bhandari:2021thi,Petroff:2021wug,Xiao:2021omr,Zhang:2022uzl,Wu:2024iyu,Glowacki:2024cgu} for recent reviews.

However, FRB cosmology based on DM is seriously hindered by parameter degeneracies, which result in two main sources of systematic uncertainties \cite{Kumar:2019qhc}.
The first is from extracting the IGM contribution ($\mathrm{DM}_{\rm IGM}$) from the total DM budget (see Eq.~(\ref{DM budgt})). This is mainly due to the difficulty in modeling both the host contribution ($\mathrm{DM}_{\rm host}$) and the LoS inhomogeneity in the IGM. 
Macquart20 modeled their probability distributions and Refs.~\cite{Zhang:2020mgq,Zhang:2020xoc} further employed the IllustrisTNG (TNG) simulations to derive priors on these empirical model parameters for different FRB populations, which could help mitigate biases from host diversity and IGM fluctuations.
However, the second, and more fundamental, uncertainty from the analysis of the extracted $\mathrm{DM}_{\rm IGM}$ remains challenging to control. This is because its redshift relation (known as the Macquart relation) exhibits intrinsic degeneracies among $\Omega_{\rm b}$, $H_0$, and the baryon fraction in diffuse ionized gas ($f_{\rm d}$) (see Eq.~(\ref{aveDM})).
Given the limited constraints on $f_{\rm d}$, it is commonly modeled as either a constant or a redshift-dependent function in the literature \cite{Jaroszynski:2018vgh,Li:2020qei,Qiang:2020vta,Zhao:2020ole,Qiu:2021cww,Wang:2022ami,gjz2023,Liu:2025fdf}.
On the other hand, $H_0$ is coupled to $\Omega_{\rm b}$ through the critical density of the Universe.
Although $H_0$ has been precisely measured, estimates from early- (e.g., Planck 2018 CMB \cite{2020Planck} (hereafter Planck18)) and late-Universe (e.g., SH0ES Cepheid-calibrated Type Ia supernovae (SNe) \cite{Riess:2021jrx}) observations show a significant discrepancy now exceeding $5\sigma$, which has widely been known as the ``$H_0$ tension'' 
\cite{Guo:2018ans,DiValentino:2021izs,Abdalla:2022yfr,Hu:2023jqc,CosmoVerseNetwork:2025alb}. This inconsistency would introduce systematic uncertainties into the baryon census using FRB samples.
It should be noted that Macquart20 and Yang22 derived their respective estimates of $\Omega_{\rm b}$ by assuming $H_0 = 70$ \kms and a flat prior over $67$--$74$ \kms, which could unavoidably bias the baryon estimate in the context of the Hubble tension and thereby motivate a direct inference of $\Omega_{\rm b}$, e.g., jointly analyzing FRBs and datasets that independently measure $H_0$ in the late Universe.

As an emerging cosmological probe, gravitational-wave (GW) standard sirens offer unique advantages for measuring $H_0$ \cite{Schutz:1986gp}. This primarily arises from the fact that the amplitudes of GW waveforms are inversely proportional to the luminosity distances ($D_{\rm L}$) to the sources, allowing for a direct measurement of $D_{\rm L}$ through waveform analysis. Since $D_{\rm L} \propto 1/H_0$, GW standard sirens serve as a powerful and self-calibrating probe for determining $H_0$, provided that the redshifts of the GW sources are also known \cite{Dalal:2006qt,Cutler:2009qv,Zhao:2010sz,Cai:2016sby,LIGOScientific:2017adf,LIGOScientific:2017vwq,LIGOScientific:2017zic,Chen:2017rfc,Cai:2017aea,Wang:2018lun,Zhang:2019loq,Zhang:2018byx,Zhang:2019ylr,Wang:2019tto,Zhao:2019gyk,Gray:2019ksv,Jin:2020hmc,Yu:2020vyy,Mastrogiovanni:2021wsd,KAGRA:2021duu,Wang:2021srv,Zhu:2021aat,Zhu:2021bpp,Wu:2022dgy,Muttoni:2023prw,Yun:2023ygz,Yu:2023ico,Han:2023exn,Feng:2024lzh,Song:2022siz,Jin:2023sfc,Jin:2023tou,Li:2023gtu,Dong:2024bvw,Xiao:2024nmi,Zhu:2024qpp,Zheng:2024mbo,Han:2025fii,Du:2025odq,Song:2025bio,Jin:2025dvf,Dong:2025ikq}. 
During the first three observing runs of the LIGO, Virgo, and KAGRA (LVK) collaboration, approximately 100 GW events were detected and compiled into the GWTC-3 catalog. Using 47 GW standard sirens from GWTC-3, Ref.~\cite{LIGOScientific:2021aug} achieved a $\sim 10\%$ precision on $H_0$, and Ref.~\cite{Song:2025ddm} further combined these sirens with an additional strong-lensing system to obtain a model-independent $\sim 8\%$ constraint. {More recently, with the release of the LVK O4a data set, over 300 GW events have been reported, and by analyzing 142 of them while varying the population-model parameters of both binary black holes (BBHs) and binary neutron stars (BNSs), the LVK collaboration has obtained a $\sim 14\%$ constraint on $H_0$ \cite{LIGOScientific:2025jau}.} Therefore, the $H_0$ information uniquely provided by GW data enables a direct estimate of $\Omega_{\rm b}$ disentangled from $H_0$ in the Macquart relation, which offers new insight into resolving systematic effects in FRB cosmology. 

In this Letter, we present the first $H_0$-free baryon census at low redshift, within a late-Universe probe framework that unifies $104$ localized FRBs and $47$ GW events from the GWTC-3 catalog to jointly constrain $\Omega_{\rm b}$ and $H_0$.
Note that we do not assume any FRB/GW associations (see Refs.~\cite{Wei:2018cgd,Qiang:2024erm,wtv1-dmrw} for studies based on such associations).



\paragraph{Localized FRBs.} \label{dm}
The observed dispersion measure can be decomposed into several components:
\begin{equation}\label{DM budgt}
\rm{DM}_{\rm{obs}}=\rm{DM}_{\rm{MW,ISM}}+\rm{DM}_{\rm{MW,halo}}+\rm{DM}_{\rm{IGM}}+\rm{DM}_{\rm{host}},
\end{equation}
where the Milky Way contribution includes the interstellar medium (ISM) contribution $\mathrm{DM}_{\mathrm{MW,ISM}}$ estimated from the electron density models such as NE2001 \cite{Cordes:2002wz}, based on the source coordinates.
The halo contribution of the Milky Way ($\rm DM_{\rm{MW,halo}}$) is estimated to lie in the range $50$--$80~\rm{pc~cm^{-3}}$ \cite{Prochaskahalo}. Thus, we assume a Gaussian distribution $P_{\rm{halo}}(\mathrm{DM}_{\rm{halo}})$ with the mean value of $\mu_{\rm{halo}}= 65~\rm{pc~cm^{-3}}$ and standard deviation of $\sigma_{\rm{halo}}= 15~\rm{pc~cm^{-3}}$ \cite{Yang:2022ftm}, i.e.,
\begin{equation}\label{pdf:halo}
P_{\rm{halo}}(\mathrm{DM}_{\rm{halo}}) = 
\frac{1}{\sqrt{2\pi}\sigma_{\rm{halo}}}
\exp\left[
-\frac{(\mathrm{DM}_{\rm{halo}}- \mu_{\rm{halo}})^2}{2\sigma_{\rm{halo}}^2}
\right].
\end{equation}

The average $\mathrm{DM}_{\mathrm{IGM}}$ in the standard $\Lambda$CDM (cosmological constant $\Lambda$ + cold dark matter) cosmology is given by the Macquart relation,
\begin{align}\label{aveDM}
\langle\mathrm{DM}_{\mathrm{IGM}}\rangle
=&\frac{3cf_{\rm e}}{8 \pi G m_{\mathrm{p}}} \Omega_{\mathrm{b}}H_0 f_{\mathrm{d}} \int_0^z \frac{(1+z) dz}{\sqrt{\Omega_{\rm m}(1+z)^3 + (1-\Omega_{\rm m})}},
\end{align}
where $\Omega_{\rm m}$ is the present matter density parameter, $c$ is the speed of light, $G$ is the gravitational constant, $m_{\rm{p}}$ is the proton mass, and $f_{\rm e}$ is the number of ionized electrons per baryon, which can be approximated as $7/8$ for $z < 3$.

The IGM plasma is inhomogeneous, and the true value of $\rm DM_{\rm IGM}$ would vary significantly around the mean value $\langle{\rm{DM}}_{{{\rm{IGM}}}}\rangle$ due to the LoS variance caused by intersecting foreground galaxy halos. Thus, the probability distribution of $\rm DM_{\rm IGM}$ is
\begin{equation}\label{pdf:igm}
	P_{\rm{IGM}}(\Delta)=A \Delta^{-\beta} \exp \left[-\frac{\left(\Delta^{-\alpha}-C_0\right)^2}{2 \alpha^2 \sigma_{\rm{IGM}}^2}\right],~\Delta>0,
\end{equation}
where $\Delta \equiv {\rm{DM}}_{\rm{IGM}}/\left\langle{\rm{DM}}_{{{\rm{IGM}}}}\right\rangle$. $\alpha$ and $\beta$, which describe the inner gas density profile of halos, are best fitted as $\alpha=3$ and $\beta=3$ according to hydrodynamic simulations. The effective standard deviation  $\sigma_{\mathrm{IGM}}=F z^{-1/2}$ is parameterized by Macquart20, where $F$ is the parameter reflecting the galactic feedback. Here, $A$ is a normalization constant, and $C_0$ is chosen to ensure that the mean of this distribution is unity. 
Note that we include the factor $1/\langle{\rm DM_{\rm IGM}}\rangle$ suggested by Ref.~ \cite{Zhang:2025wif} when normalizing Eq.~(\ref{pdf:igm}).

The distribution of $\rm{DM}_{\rm{host}}$ is modeled as
\begin{align}\label{pdf:host}
P_{\rm{host}}(\rm{DM}_{\rm{host}}) 
&= \frac{1}{\sqrt{2\pi}\sigma_{\rm{host}}\mathrm{DM}_{\rm{host}}} \nonumber \\
&\times \exp\left[
-\frac{(\ln \mathrm{DM}_{\rm{host}}- \mu_{\rm{host}})^2}{2\sigma_{\rm{host}}^2}
\right],
\end{align}
where a correction $\mathrm{DM}_{\text {host }} \rightarrow \mathrm{DM}_{\text {host}} /(1+z)$ has been applied.
The mean value and variance of the distribution are ${\rm e}^{\mu_{\rm host}}$ and ${\rm e}^{2\mu_{\rm host}+\sigma_{\rm{host}}^2}({\rm e}^{\sigma_{\rm{host}}^2}-1)$, respectively.

Thus, the total likelihood of the localized FRB data set $\{x_{\rm FRB}\}$ composed of $N_{\rm FRB}$ sources, is given by
\begin{equation}\label{eq:FRB likelihood}
\mathcal{L}_{\mathrm{FRB}}(\{x_{\rm FRB}\}) = \sum_{i=1}^{N_{\mathrm{FRB}}} P_i\left(\mathrm{DM}'_i\right).
\end{equation}
For the $i$-th FRB, the possibility distribution of its extragalactic DM component ($\mathrm{DM}' = \rm DM_{\rm IGM}+\rm DM_{\rm host}$) is computed by convolving Eqs.~(\ref{pdf:halo}), (\ref{pdf:igm}), and~(\ref{pdf:host}):
\begin{align}
P_i\left(\mathrm{DM}'_i\right) 
&=  \int_{30\,\mathrm{pc\,cm}^{-3}}^{100\,\mathrm{pc\,cm}^{-3}} 
    P_{\rm halo}\left(\mathrm{DM}_{\rm halo}\right) \,\mathrm{dDM}_{\rm halo} \nonumber \\
& \times \int_{0}^{\mathrm{DM}'_i - \mathrm{DM}_{\rm halo}} 
    P_{\rm host}\left(\mathrm{DM}_{\rm host}\right) \nonumber \\
& \times P_{\rm IGM}\left(\mathrm{DM}'_i - \mathrm{DM}_{\rm halo} - \mathrm{DM}_{\rm host} \right)
    \,\mathrm{dDM}_{\rm host}.
\end{align}


Finally, we select a sample of $104$ localized FRBs (see the Supplemental Material~\cite{SM} (see also Refs.~\cite{Chatterjee:2017dqg,Tendulkar2016,Bhandari2022,Bannister2019,Prochaska2019,Bhardwaj2024,Bhandari2020,Ibik2024,Michilli2023,Ravi2019,Chittidi2021,Heintz2020,Law2020,Shin:2024ezg,Ravi2022,Shannon2024,Caleb2023,Cassanelli2023,Gordon2023,Sharma2024,Law2024,Li:2025ckl,Ryder2023,Rajwade2024,AnnaThomas2025,2025MNRAS.538.1800H,connor2025gas,2025ApJ...979L..22E,Amiri:2025sbi,Tian:2024ygd} therein).
The dataset is selected from an initial set of $119$, based on the following criteria:
(1) $\mathrm{DM}_{\mathrm{MW,ISM}}$ is required to be less than $40\%$ of $\mathrm{DM}_{\rm obs}$ \cite{connor2025gas};
(2) $\mathrm{DM}_{\rm obs} - \rm DM_{MW,ISM}$ is taken to exceed 80~\pccc, ensuring a positive $\mathrm{DM}_{\rm host}$;
(3) FRB~20190520B and FRB~20220831A are excluded due to excess local DM contributions from their host environments.

\paragraph{GW standard sirens.}
We use 47 GW standard sirens with SNRs greater than 11 and an inverse false alarm rate exceeding 4 yr from GWTC-3\footnote{\url{https://gwosc.org/GWTC-3}}, covering a redshift range of approximately $z < 0.8$. The sample consists of 42 BBHs, three neutron star-black holes (NSBHs), and two BNSs. We do not consider GW events detected in the LVK O4a run, mainly because during this run, Virgo is not online, and all GW events are detected by two LIGOs, resulting in relatively poorer sky localizations and luminosity distance measurements \cite{LIGOScientific:2025jau}.

We obtain the redshift of GW170817 from the electromagnetic (EM) counterpart \cite{LIGOScientific:2017adf,LIGOScientific:2017vwq,LIGOScientific:2017zic}. For GW events without EM counterparts, we infer their redshifts through the dark siren method \cite{Schutz:1986gp,Mastrogiovanni:2023emh,Gray:2023wgj} and incorporate the GLADE+ catalog \cite{Dalya:2021ewn}. In the dark siren method, assuming the GW sources all originate from galaxies, we can cross-match sky localizations of GW events with galaxies collected in galaxy catalogs, and identify a serious potential host galaxy. These potential host galaxies provide redshift priors for GW events, which help to infer cosmological parameters in a hierarchical Bayesian framework. Furthermore, GW population models can help break degeneracies of masses and redshifts in GW detections, thereby providing additional redshift information \cite{Taylor:2011fs,Ezquiaga:2022zkx,Mastrogiovanni:2023emh,Gray:2023wgj}. 
{In this paper, we use the public pipeline \texttt{ICAROGW} \cite{Mastrogiovanni:2023zbw} to compute the GW likelihood function, denoted as $\mathcal{L}_{\rm GW}$. The explicit formulas can be found in Refs~\cite{Mastrogiovanni:2023emh,Gray:2023wgj}.}

{We adopt the same population model assumptions as Ref.~\cite{LIGOScientific:2021aug}, but restrict our analysis to the Power Law + Peak BBH mass model, which is best supported by the GWTC-3 data. Considering additional models would increase computational cost and lies beyond the scope of this work. For BNSs, we assume that neutron star masses are uniformly distributed in the range of $1-3$ $M_{\odot}$. In the case of NSBH systems, the black hole component is drawn from the same distribution as the primary black holes in BBHs, while the neutron star component follows the mass distribution assumed for BNSs. The redshift distribution of GW sources is modeled using a phenomenological prescription that traces the cosmic star formation rate. Consistent with Ref.~\cite{LIGOScientific:2021aug}, we disregard potential redshift evolution in the BBH mass distribution as well as variations in the GW detection rate arising from the spin distribution. These effects are currently subdominant compared to the statistical uncertainties present in the existing GW observations.}


\paragraph{Joint analysis of FRB and GW data.}

We jointly constrain cosmological and all DM parameters with flat priors for the following seven parameters: 
$\Omega_{\rm m} \in \mathcal{U}(0.28, 0.35)$ \cite{Yang:2022ftm}, 
$\Omega_{\rm b} \in \mathcal{U}(0.01, 0.10)$, 
$H_{0} \in \mathcal{U}(20, 140)~\rm km\ s^{-1}\ Mpc^{-1}$, 
$f_{\rm d} \in \mathcal{U}(0.84, 0.96)$,\footnote{This prior on $f_{\rm d}$ is informed by observations of the baryon fraction in stars ($f_*$) and cold gas ($f_{\rm cold}$), independent of FRBs. Assuming a conservative flat prior for $f_*$ extended by a factor of five, the diffuse baryon fraction is inferred via $f_{\rm d} = 1 - f_* - f_{\rm cold}$ \cite{connor2025gas}.} 
$F \in \mathcal{U}(0.011, 1.0)$, 
$\mu_{\rm host} \in \mathcal{U}(3.0, 5.3)$, 
and $\sigma_{\rm host} \in \mathcal{U}(0.2, 2.0)$ \cite{Macquart:2020lln}. 
To test the robustness of our analysis, we examine the effect of prior assumptions on our results.  
For the $\Omega_{\rm m}$ prior, we relax it to a broad range $\mathcal{U}(0,1)$ to assess its impact on the constraints. 
For the GW population-model parameters, following Ref.~\cite{LIGOScientific:2021aug}, we first fix all parameters to their fiducial values and then, vary the BBH mass- and redshift-distribution parameters of GW sources, adopting the same prior ranges as in the spectral-siren analysis of Ref.~\cite{LIGOScientific:2021aug}, to assess the impact of population-model uncertainties on our results. This treatment is similar to Ref.~\cite{Gray:2023wgj}; however, unlike Ref.~\cite{LIGOScientific:2025jau}, we do not free the mass-distribution parameters of neutron-star sources.

We utilize the {\tt bilby}\footnote{\url{https://github.com/bilby-dev/bilby}} package to perform Markov Chain Monte Carlo (MCMC) analysis and choose the {\tt emcee}\footnote{\url{https://emcee.readthedocs.io/en/stable}} sampler. The {\tt GetDist}\footnote{\url{https://github.com/cmbant/getdist}} package to plot the marginalized posterior distributions. The final likelihood $\mathcal{L}$ used in our analysis is the combination of $\mathcal{L}_{\mathrm{FRB}}$ and $\mathcal{L}_{\rm GW}$, which is written as 
\begin{equation}
\ln \mathcal{L}=\ln \left(\mathcal{L}_{\mathrm{FRB}}\right)+\ln \left(\mathcal{L}_{\mathrm{GW}}\right).
\label{Eq:Eq11}
\end{equation}

\begin{table}[htbp]
\caption{$1\sigma$ constraints on cosmological parameters \texorpdfstring{$\Omega_{\rm b}$}{Omega_b} and \texorpdfstring{$H_0$}{H0}. The results are derived from $104$ localized FRBs alone or in combination with $47$ GW events, under different assumptions. {Unless otherwise specified, the FRB + GW case assumes a flat prior of $\Omega_{\rm m} \in \mathcal{U}(0.28,0.35)$ and fixed GW population parameters.} Note that $H_0$ is in units of \kms.}
\label{tab:1}
\setlength{\tabcolsep}{0.3mm}
\renewcommand{\arraystretch}{1.5}
\begin{center}
\begin{tabular}{lcc}
\hline \hline
Data & $\Omega_{\rm b}$ & $H_0$ \\
\hline
FRB & $0.0520^{+0.0120}_{-0.0280}$ & $77.0^{+20.0}_{-40.0}$ \\
FRB + $H_0$ (Planck18 prior) & $0.0513\pm 0.0046$ & $67.4\pm 0.5$ \\
FRB + $H_0$ (SH0ES prior) & $0.0473\pm0.0043$ & $73.0\pm 1.0$ \\
FRB + $\Omega_{\rm b} h^2$ (BBN prior) & --- & $66.2^{+5.3}_{-7.3}$ \\ 
\hline
FRB + GW & $0.0488\pm0.0064$ & $71.6^{+4.4}_{-8.6}$ \\
{FRB + GW (GW parameters free)} & $0.0448^{+0.0094}_{-0.0076}$ & $79.9^{+5.0}_{-17.0}$ \\
FRB + GW ($\Omega_{\rm m} \in \mathcal{U}(0,1)$) & $0.0454^{+0.0066}_{-0.0082}$ & $72.7^{+4.7}_{-9.0}$ \\
\hline \hline
\end{tabular}
\end{center}
\end{table}



\paragraph{Results.}\label{sec3.1}

\begin{figure}[htbp]
\centering 
\includegraphics[width=\linewidth, angle=0]{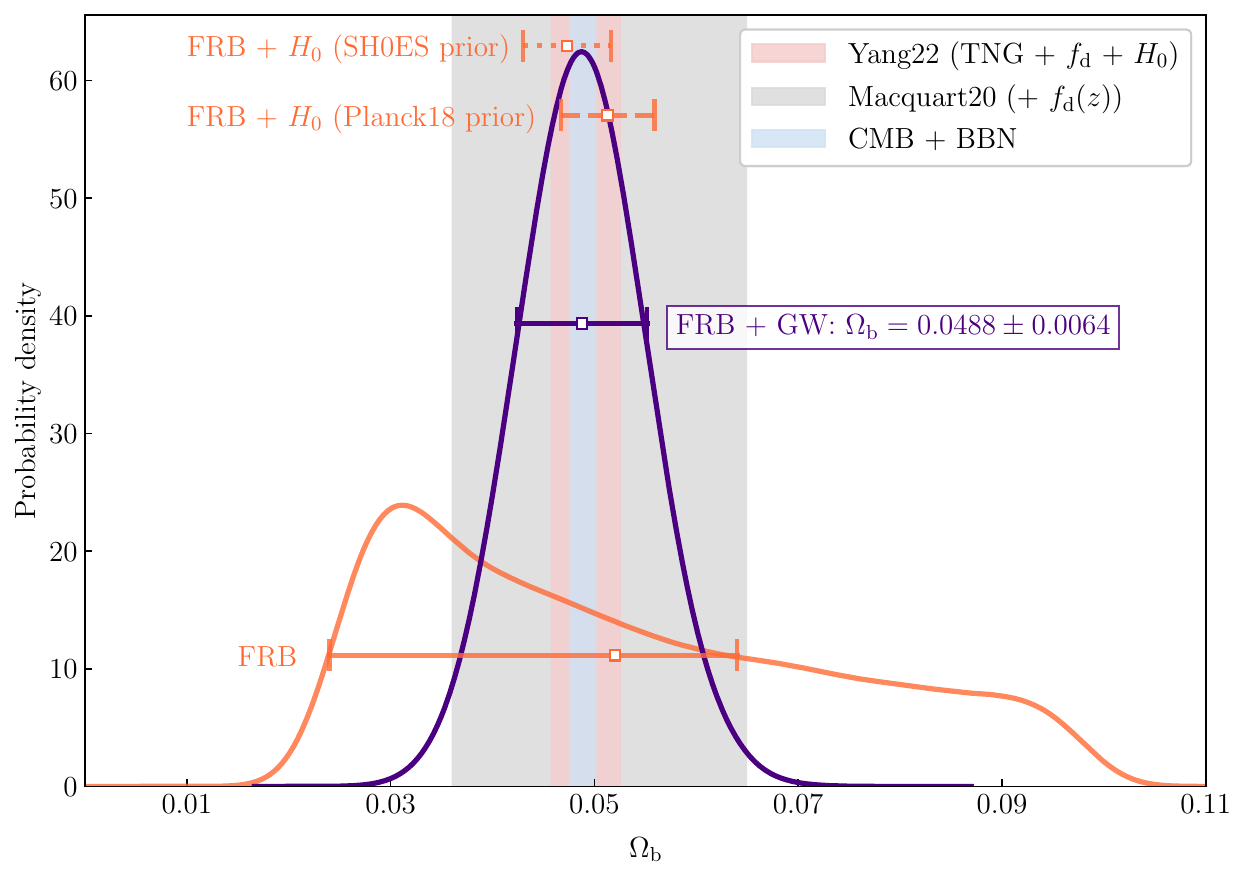}
\vspace{-0.5cm} 
\caption{
Posteriors for \ob from 104 FRBs (orange) and 104 FRBs + 47 GWs (indigo). The joint constraint is $\Omega_{\rm b} = 0.0488 \pm 0.0064$, obtained by marginalized over all DM parameters. 
Errorbars in different colors and linestyles are shown under different assumptions: FRB + \(H_0\) (Planck18 prior) and FRB + \(H_0\) (SH0ES prior) adopt Gaussian priors on \(H_0\) from Planck18~\cite{2020Planck} and SH0ES~\cite{Riess:2021jrx}, respectively; 
Shaded bands in different colors show the results from CMB + BBN \cite{Cooke:2017cwo}, Macquart20, and Yang22 \cite{Yang:2022ftm}.
Note that Yang22 adopts the prior on DM model parameters from the TNG simulation \cite{Zhang:2020mgq,Zhang:2020xoc}, a flat prior on \(H_0\) \(\in\) \(\mathcal{U}(67, 73)\)~km\,s\(^{-1}\)\,Mpc\(^{-1}\), and $f_{\rm d}=0.82$. All results are at the $1\sigma$ level.
}
\label{Fig:1}
\end{figure}

\begin{figure}[htbp]
\includegraphics[width=\linewidth, angle=0]{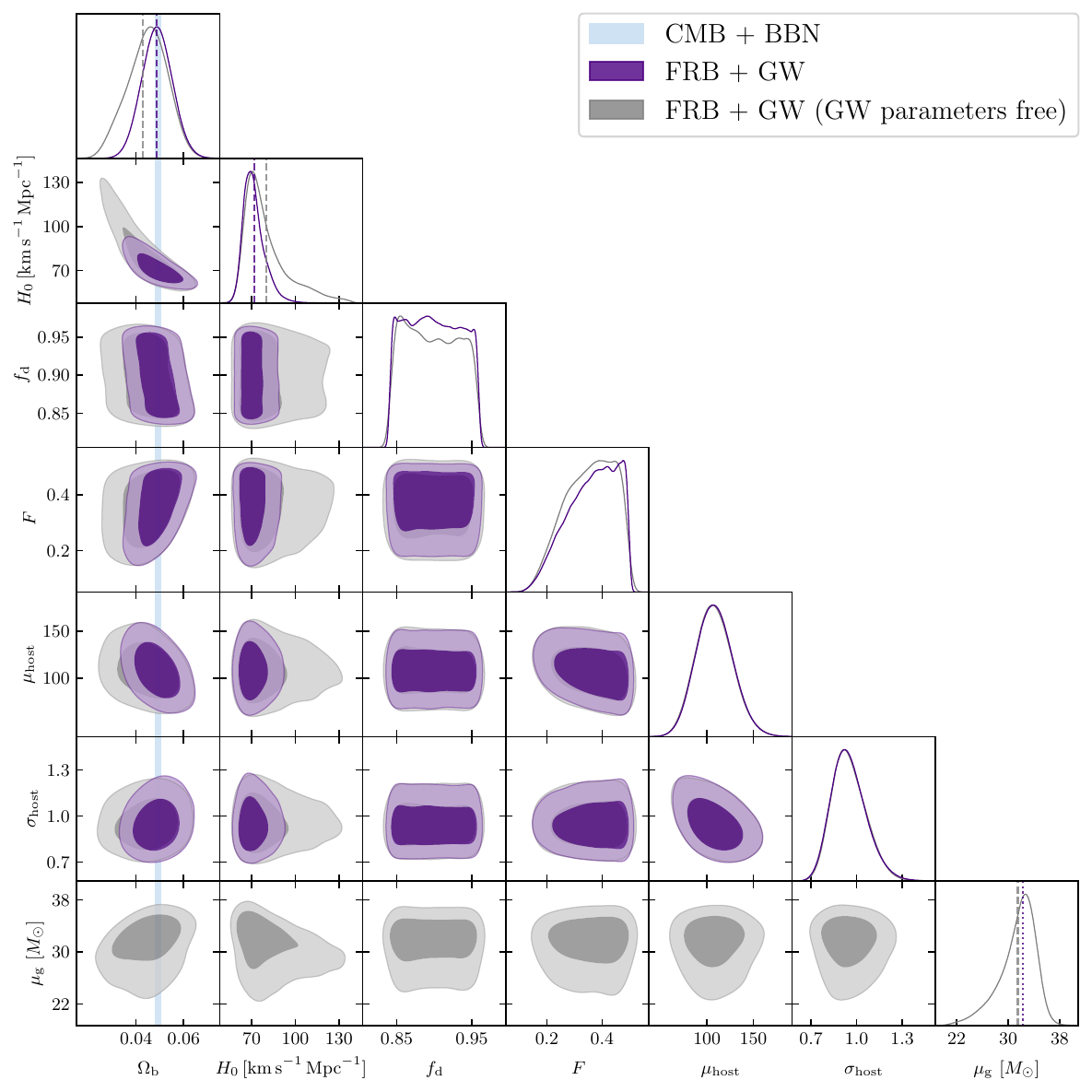}
\vspace{-0.5cm} 
\caption{
{Posteriors ($1\sigma$ and $2\sigma$ credible regions) for $\Omega_{\rm m}$, $H_0$, and the FRB DM parameter $f_{\rm d}$, $F$, $\sigma_{\rm host}$, $\mu_{\rm host}$, and the typical GW population parameter $\mu_{\rm g}$ from FRB + GW data. Two cases are shown: GW parameters fixed within the PowerLaw + Peak model (indigo) and the parameters allowed to vary freely and marginalized over (gray). 
The dotted lines indicate the median values of each posterior, except for $\mu_{\rm g}$ in the GW-parameter-fixed case, where the dotted line marks the prior value $\mu_{\rm g} = 32.27~M_\odot$ adopted following Ref.~\cite{LIGOScientific:2021aug}.}
}
\label{Fig:2}
\end{figure}

We report the main results for the baryon-density estimate in the first block of Table~\ref{tab:1}.
The posteriors for \ob\ and the corresponding credible intervals are shown in Fig.~\ref{Fig:1}.
Figure~\ref{Fig:2} shows the joint posterior distributions of the cosmological and DM parameters.
Using FRB data alone yields a relatively weak constraint of $\Omega_{\mathrm{b}}=0.052_{-0.028}^{+0.012}$ at the $1\sigma$ level.
Incorporating GW standard sirens significantly narrows this to $\Omega_{\mathrm{b}}=0.0488\pm0.0064$, by breaking the strong degeneracy between $\Omega_{\mathrm{b}}$ and $H_0$ via the independent $H_0$ measurement provided by GWs. The $\Omega_{\mathrm{b}}$-$H_0$ panel in Fig.~\ref{Fig:2} shows that this correlation is markedly stronger than those between $\Omega_{\mathrm{b}}$ and the other parameters. This late-Universe estimate exhibits remarkable concordance with early-Universe values from BBN and CMB at the $\sim13\%$ level.
For the FRB DM parameters, we obtain $f_{\rm d}=0.90\pm0.04$, $F=0.36^{+0.13}_{-0.05}$, $\mu_{\rm host}=4.67^{+0.20}_{-0.16}$, and $\sigma_{\rm host}=0.95^{+0.09}_{-0.13}$, corresponding to a median rest-frame host DM of $108\pm20~\mathrm{pc~cm^{-3}}$ and a mean host DM of $168^{+57}_{-40}~\mathrm{pc~cm^{-3}}$, consistent with previous estimates \cite{Yang:2024vqq,connor2025gas}.

We compare our \ob results with those from previous works. Macquart20 and Yang22 constrained \ob using different samples of localized FRBs under external assumptions of $H_0 = 70$ \kms and a flat prior on $H_0 \in \mathcal{U}(67,73)$ \kms, as shown by the gray and red bands in Fig.~\ref{Fig:1}, respectively.
The constraint from Yang22 compares favorably to ours, since the current GW standard siren sample remains too limited to outperform their adopted $H_0$ prior.
Applying the external $H_0$ priors from Planck18 or SH0ES to our FRB sample alone yields $\Omega_{\rm b} = 0.0513\pm0.0046$ and $\Omega_{\rm b}=0.0473\pm0.0043$, respectively; the corresponding error bars are shown in Fig.~\ref{Fig:1}.
Obviously, the $H_0$ tension would lead to two biased \ob estimates.
Although the induced ``$\Omega_{\rm b}$ inconsistency'' in our sample is within $1\sigma$ uncertainty ($\sim 0.95\sigma$), it may become non-negligible with larger samples in updated DM analyses. 
Nevertheless, the joint FRB + GW result assumes no prior on $H_0$ and instead provides a late-Universe measurement of baryons, independent of both CMB and the local distance ladder.  
This FRB + GW synergy may establish a novel paradigm for the cosmic baryon census, analogous to CMB + BBN, and serve as a powerful low-redshift cosmological probe in the era of next-generation FRB and GW surveys \cite{Zhang:2023gye}.


We assess the impact of GW population modeling on our baryon census.
In our main analysis, no FRB population model is imposed, whereas the PowerLaw + Peak GW population model is adopted for the GW population, as is commonly used in GW-based cosmological inference \cite{LIGOScientific:2021aug}. To investigate the robustness of our \ob estimate against this assumption, we release the GW population model parameters and marginalize over them.
We find that joint FRB + GW analysis with GW model parameters free yields a higher value of $H_0 = 79.9^{+5.0}_{-17.0}$ \kms and a correspondingly lower value of $\Omega_{\rm b} = 0.0448^{+0.0094}_{-0.0076}$. The $\Omega_{\rm b}$ result ($\sim 20\%$ precision) remains broadly consistent with the CMB + BBN determination at the $1\sigma$ level.

To illustrate how the GW population prior affects $\Omega_{\rm b}$, we focus on the parameter $\mu_{\rm g}$, which denotes the mean of the Gaussian component in the primary mass distribution within the PowerLaw + Peak model. This parameter is closely correlated with the inferred $H_0$ from GW observations \cite{LIGOScientific:2021aug}, making it suitable for accounting for the impact of population modeling.
As shown in Fig.~\ref{Fig:2}, $\mu_{\rm g}$ exhibits a negative degeneracy with $H_0$ and a corresponding positive degeneracy with $\Omega_{\rm b}$.
Consequently, compared with the prior value $\mu_{\rm g} = 32.27~M_\odot$ adopted in the GW-parameter-fixed (indigo) case, the lower inferred value of $\mu_{\rm g} = 31.5~M_\odot$ in the GW-parameter-free (gray) case leads to a higher $H_0$ and thus a lower $\Omega_{\rm b}$.
Thus, to obtain a robust estimate of $\Omega_{\rm b}$, caution is required when adopting GW-population assumptions, due to their strong impact on the inferred $H_0$. Nevertheless, our test without imposing any GW population model reinforces that our fiducial result obtained with the PowerLaw + Peak model is reliable and not significantly biased.

We also assess the dependence on the prior for $\Omega_{\rm m}$. Relaxing the prior from $\Omega_{\rm m}\in\mathcal{U}(0.28,0.35)$ to a broad flat prior $\Omega_{\rm m}\in\mathcal{U}(0,1)$ yields $\Omega_{\rm b}=0.0454^{+0.0066}_{-0.0082}$ and $H_0=72.7^{+4.7}_{-9.0}$ \kms. Compared with the fiducial case ($\Omega_{\rm b}=0.0488\pm0.0064$), the slightly lower value of $\Omega_{\rm b}$ is driven by the biased constraint on $\Omega_{\rm m}$ under the broad prior, $\Omega_{\rm m}=0.225^{+0.060}_{-0.180}$, which propagates through the positive correlation between \ob and $\Omega_{\rm m}$. Therefore, we caveat that our present analysis moderately depends on the $\Omega_{\rm m}$ prior, which can be improved with larger future FRB and GW samples to better constrain $\Omega_{\rm m}$.

\begin{figure}[!tbp]
\centering
\includegraphics[width=\linewidth, angle=0]{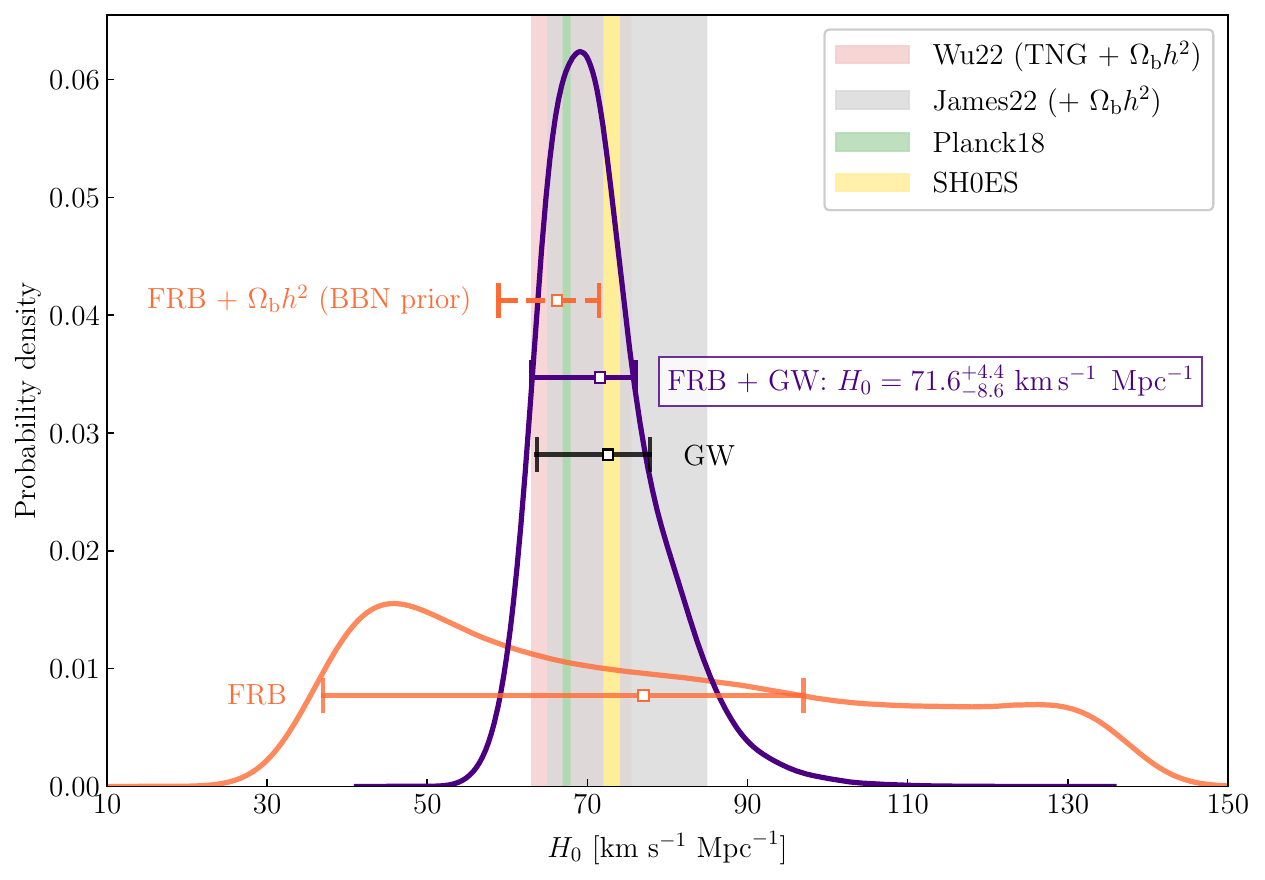}
\vspace{-0.5cm} 
\caption{
Same as Fig.~\ref{Fig:1} but for posteriors for $H_0$. 
The joint FRB + GW constraint is $H_0 = 71.6^{+4.4}_{-8.6}$ km\,s$^{-1}$\, Mpc$^{-1}$.
The orange and black errorbars correspond to constraints from FRB + $\Omega_{\rm b}h^2$ (BBN prior~\cite{Cooke:2017cwo}) and GW data alone, respectively. 
The green, yellow, red, and gray shaded bands denote the results from Planck18 \cite{2020Planck}, SH0ES \cite{Riess:2021jrx}, Refs.~\cite{Wu:2021jyk} (Wu22), and \cite{James:2022dcx} (James22), respectively. Note that Wu22 adopts the TNG prior, and both Wu22 and James22 adopt $\Omega_{\rm b}h^2$ priors from BBN \cite{Cooke:2017cwo} and CMB \cite{2020Planck}, respectively. All results are at $1\sigma$ uncertainty.
}
\label{Fig:3}
\end{figure}

Last, the corresponding constraints on $H_0$ are also listed in Table~\ref{tab:1}, with the posteriors and credible intervals shown in Fig.~\ref{Fig:3}. A direct measurement of $H_0$ using FRBs alone is difficult due to strong parameter degeneracies. Concretely, our FRB-only measurement gives $H_0 = 77.0^{+20.0}_{-40.0}$ \kms, while including GW data narrows the constraint to $H_0 = 71.6^{+4.4}_{-8.6}$ \kms. This improvement is primarily driven by the GW dataset itself, which yields $H_0 = 72.6^{+5.2}_{-8.9}$ \kms.
Previous FRB studies commonly inferred $H_0$ by fixing the baryon density to early-Universe values.
For example, Refs.~\cite{Wu:2021jyk} and \cite{James:2022dcx} inferred $H_0$ from different FRB samples by adopting external priors on
$\Omega_{\rm b}h^2$, namely 
$\Omega_{\rm b}h^2 \sim \mathcal{N}(0.02235,0.00049)$ from BBN and 
$\Omega_{\rm b}h^2 = 0.02242$ from Planck18, respectively, as illustrated by the red and gray bands in Fig.~\ref{Fig:3}.
Using the same BBN prior, we obtain $H_{0}=66.2^{+5.3}_{-7.3}$ \kms\ from FRB + $\Omega_{\rm b}h^{2}$, with a precision comparable to that of FRB + GW.
Unlike these prior-based methods, the FRB + GW constraint provides a late-Universe determination of $H_0$.
More data is required to offer further insight into arbitrating the $H_0$ tension.

\paragraph{Conclusion.}
In this Letter, we combine two emerging cosmological probes, i.e., $104$ localized FRBs and $47$ GW standard sirens, within a unified framework to measure the cosmic baryon density in a $H_0$-free approach. 
Previous baryon censuses often fixed \(H_0\) or used external priors, which could bias the inference under the Hubble tension.  
If FRB DMs are employed to trace the baryons, serious parameter degeneracies arise in the DM modeling, particularly between $H_0$ and $\Omega_{\rm b}$.
GWs independently infer $H_0$ from self-calibrated luminosity distances obtained through waveform analysis as sirens, thus breaking the $\Omega_{\rm b}$-$H_0$ degeneracy.
This joint analysis yields a constraint of $\Omega_{\rm b} = 0.0488 \pm 0.0064$ ($1\sigma$), which is remarkably consistent with early-Universe values from BBN and CMB at the $\sim13\%$ level and provides a purely late-Universe perspective on resolving the missing baryon problem.
We also obtain a simultaneous constraint on the Hubble constant, $H_0 = 71.6^{+4.4}_{-8.6}~\mathrm{km\,s^{-1}\,Mpc^{-1}}$, primarily driven by the GW data. Together, these results exhibit $\Omega_{\rm b}$-$H_0$ concordance between the early and late Universe.

We note that our baryon census still depends on the adopted $\Omega_{\rm m}$ prior and the assumed GW population model. Nevertheless, the $\Omega_{\rm b}$ values inferred when removing these assumptions show only modest shifts and remain consistent with CMB + BBN determinations at the $1\sigma$ level. Using larger samples of localized FRBs and GW standard sirens, future analyses would jointly constrain $\Omega_{\rm m}$, GW population parameters, and $\Omega_{\rm b}$ more robustly, thereby reducing this dependence.

Overall, the novel combination of FRBs and GWs enables a direct, unique, and robust measurement of both the baryon content and cosmological distances at low redshifts.
With an increasing number of GW detections expected in the upcoming LVK run, and potentially thousands of localized FRBs expected from CHIME/Outriggers \cite{masui2023pinpointing} and DSA-110 \cite{Law:2023ibd}, their synergy will soon emerge as a powerful low-redshift cosmological probe.

\paragraph{Acknowledgments.}
We thank Tian-Nuo Li, Meng-Lin Zhao, Yang Liu, and Qin Wu for fruitful discussions. This work was supported by the National Natural Science Foundation of China (Grants Nos. 12533001, 12575049, 12473001, and 12473091), the National SKA Program of China (Grants Nos. 2022SKA0110200 and 2022SKA0110203), the China Manned Space Program (Grant No. CMS-CSST-2025-A02), the National 111 Project (Grant No. B16009), and the Fundamental Research Funds for the Central Universities (Grant No. N2405008).

\bibliography{frbgw}

\end{document}